\documentclass[letterpaper, 10 pt, journal, twoside]{IEEEtran}
\usepackage[pdftex]{graphicx} 
\usepackage{gensymb}
\usepackage{amsmath}
\usepackage{hyperref}
\IEEEoverridecommandlockouts  
\usepackage{booktabs}
\usepackage{xcolor}
\usepackage[normalem]{ulem}
\usepackage{tikz}
\usepackage{xfrac}
\usepackage{float}
\usepackage{placeins}
\usepackage{subcaption}
\usepackage[justification=centering]{caption}
 \usepackage[pscoord]{eso-pic}
\begin{document}

\title{Analysis of Optimal Portfolio Management Using Hierarchical Clustering}
\author{
\begin{tabular}{c} Kapil Panda \\ University of North Texas \\ Denton, TX, United States of America \\ kapil.panda30sc@gmail.com \end{tabular}
}
\maketitle
\begin{abstract} Portfolio optimization is a task that investors use to determine the best allocations for their investments, and fund managers implement computational models to help guide their decisions. While one of the most common portfolio optimization models in the industry is the Markowitz Model, practitioners recognize limitations in its framework that lead to suboptimal out-of-sample performance and unrealistic allocations. In this study, I refine the Markowitz Model by incorporating machine learning to improve portfolio performance. By using a hierarchical clustering-based approach, I am able to enhance portfolio performance on a risk-adjusted basis compared to the Markowitz Model, across various market factors.
\end{abstract}

\begin{IEEEkeywords}computational finance, portfolio optimization, hierarchical clustering, machine learning, asset allocation, stocks
\end{IEEEkeywords}

\section{Introduction}
In the financial industry, portfolio optimization remains a significant point of focus among practitioners and academics alike. With portfolio optimization being a growing sector in finance over the past few years, the need for new computational methods to guide investors' asset allocations and improve accuracy has never been greater \cite{de791fb1-64f6-3d3b-9f5b-0899051b13ae}. 

One of the most common portfolio optimization models used in industry is the Mean-Variance Model, also known as the Markowitz Model, created by Harry Markowitz in 1952. Published as part of his paper ``Portfolio Selection,'' \cite{e5a1bb8f-41b7-35c6-95cd-8b366d3e99bc} Markowitz founded the Modern Portfolio Theory, a method that risk-averse investors use today to construct diversified portfolios that maximize their returns. In this theory, Markowitz states that any investment's risk and return characteristics should not be evaluated by itself but by how it affects the portfolio's overall risk and return. 

While this model is fundamentally sound and is still used today to help investors fulfill their multi-million-dollar investments, asset allocators and portfolio managers recognize its many limitations \cite{article3}. For starters, it is well known that the traditional Markowitz approach to portfolio optimization exhibits high investment turnover, poor out-of-sample performance, and odd weight allocations \cite{https://doi.org/10.1111/1540-6261.00580}. Furthermore, this model is also known to make several incorrect assumptions, such as assuming that all investors have the same credit and failing to account for additional charges (i.e., brokerage and other taxes that can sway investors' decisions) \cite{RePEc:eee:jfinec:v:33:y:1993:i:1:p:3-56}. Due to these several constraints, many practitioners are looking for new models to guide their decisions.

In recent years, advancements in machine learning have opened up new possibilities for improving portfolio allocation and optimization in the financial sector. As the demand for more accurate and efficient asset allocation methods grows, machine learning techniques have emerged as a promising solution to address the limitations of traditional portfolio optimization models.

Therefore, in this study, I aim to leverage the power of machine learning to devise a new portfolio optimization model. To improve on ex-ante portfolio optimization models, I implement a machine learning technique known as hierarchical clustering\cite{article1} to increase optimization accuracy and account for the Markowitz model's limitations. 

To do this, I devise a clustering algorithm that groups assets based on their correlations and similarities to form clusters \cite{RePEc:eee:dyncon:v:32:y:2008:i:1:p:235-258}. These clusters are then sub-clustered using recursive bisection, and then rearranged to form the covariance matrix using a Quasi-Diagonalization, in order to increase robustness in the portfolio \cite{RePEc:wsi:ijtafx:v:03:y:2000:i:03:n:s0219024900000255}. Finally, I allocate and build the portfolio by using a long-short strategy which allows me to take advantage of both upward and downward market movements.

To evaluate the model's performance, I assess the portfolio performance of the clustered model against the Markowitz Model under common model metrics\cite{258a7a1a-55ac-32d1-83ac-904915c36007} such as Average Monthly Excess
Returns, Standard Deviation, and Sharpe-Ratios to see if higher out-of-sample risk-adjusted returns are attainable.
\section{Methodology}
\subsection{Data Used}
In this research, I used the monthly returns of common stocks listed on the three major U.S. stock exchanges: NYSE, AMEX, and NASDAQ, as reported by the Center of Research in Security Prices (CRSP) from 1965-2022. To ensure the real-world implementation and practicability of the findings in this paper, I applied several filters to the data so that the stocks available for allocation are liquid and realizable for investment. Using these filters, I specifically obtained data concerned with price, return, shares out, CFACPR, and weighted value returns, for thousands of companies in various sectors. Additionally, I also obtained monthly risk-free data from Ken French’s data library that gave us market excess returns, the risk-free rate, size (SMB), value (HML), profitability (RMW), investment (CMA), and momentum (MOM) factors. In this research, I analyze the covariance matrix using three different look-back times, a 12-month, 60-month, and 120-month look-back time. 
\subsection{Algorithm}
\subsubsection{Distance Equation}
In this analysis, I use the following standard distance metric, proposed by Marcos de Prado\cite{Prado_2016}, to devise a distance matrix: 
\FloatBarrier
\begin{figure}[!htb]
	\centering
\includegraphics[width=\columnwidth]{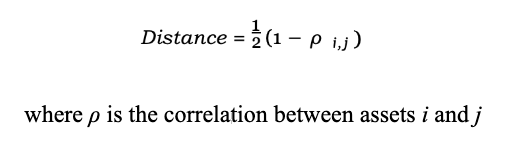}
     \captionsetup{labelformat=empty}
	\caption{Equation 1: Standard Distance Metric}
         \label{fig: Table 1}
\end{figure}
\FloatBarrier
To calculate the distances between the stocks in the cross-sections of my data, I used Equation 1 to conduct my clustering. This distance equation quantifies the pairwise distances between all data points, reflecting how closely related or distinct they are from one another.
\subsubsection{Hierarchical Clustering}
In this research, I used the hierarchical clustering machine learning approach to group stocks that are similar to one another and then placed them on a hierarchy based on their performance. 
This approach allows for a more sophisticated analysis of correlations and similarities. By grouping stocks based on their inherent relationships, the model can capture complex market dynamics and dependencies that may not be evident in traditional optimization techniques. This improves risk management by enabling the identification of coherent clusters that respond differently to market conditions, allowing investors to better understand potential sources of risk within their portfolios and make more informed decisions in varying market environments. For example, stocks can be grouped based on industries, which leads to a more robust and diversified portfolio by removing sector bias and thus resulting in improved portfolio turnover rates, which is one of the significant flaws in the Markowitz Model. 

In this framework, I start by taking each company’s monthly return values from the CRSP data and subtracting them from the risk-free rates I received from the Ken French data library to get the data’s excess return value, which is what I analyzed. Since excess return rates represent the returns achieved above the risk-free rate, which is the return an investor would earn on a risk-free asset like a government bond, by subtracting the risk-free rate from the actual returns of the portfolio, I can isolate the additional returns generated by taking on risk through investments in stocks. 

Next, to determine the correlations to use for my distance matrix, I heat-mapped the data points from various stocks to find the similarities and differences between them. Then, using the distance equation, I created a distance matrix based on the correlations between stocks in the cross-section at a given time. Finally, using the test data sets of sixty years worth of financial stock data, I clustered the stocks based on their correlations. The relationship clustering can be seen in Figure 1:
\FloatBarrier
\begin{figure}[!htb]
	\centering
	\includegraphics[width=\columnwidth]{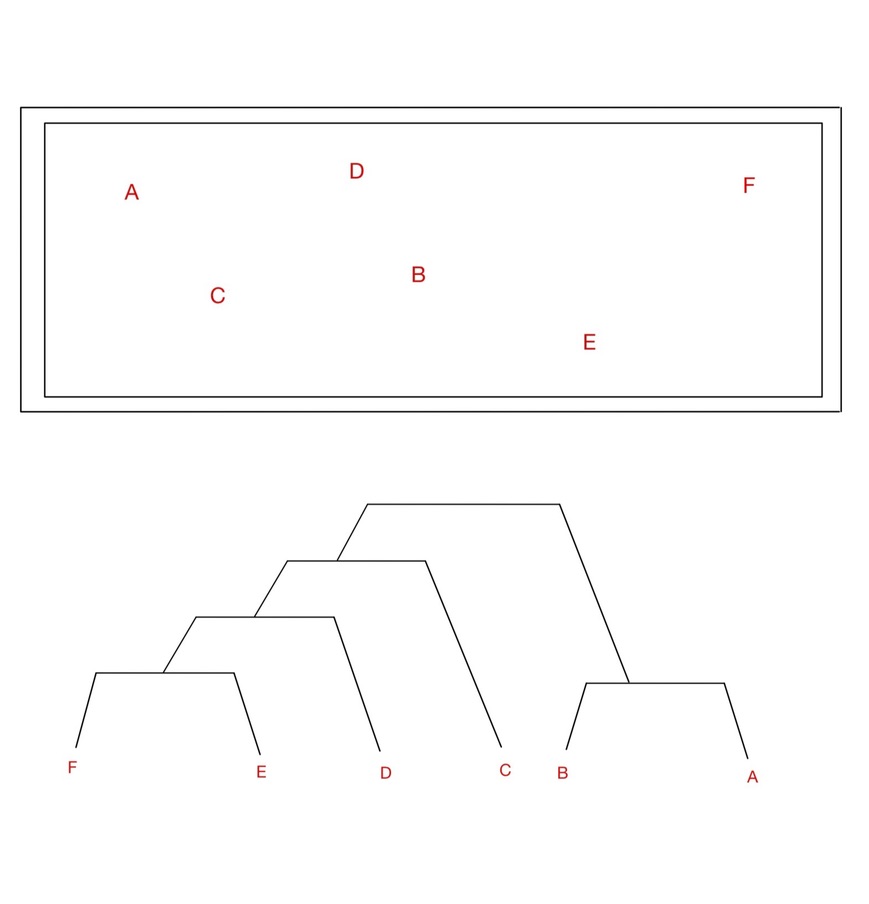}
 \captionsetup{labelformat=empty}
	\caption{Fig. 1: Hierarchical Clustering Dendogram}
         \subcaption{Figure 1 demonstrates the clustering of the stocks that the model aims to do. When given a random group of stocks, the hierarchical clustering model clusters the stocks based on their distances determined by their correlations.}
	\label{fig: Figure 1}
\end{figure}
\FloatBarrier
As seen in Figure 1, stocks A and B are first clustered together as they are the most similar, followed by stocks E and F being clustered together as well. Then, the clustering algorithm includes stock D into the clustering pair of stocks E and F and then includes stock C into the next pair. Finally, the asset pair A and B are clustered with the rest of the assets in the last step of our clustering.
\subsubsection{Recursive Bisection}
The next step is to conduct recursive bisection, which involves assigning actual portfolio weights to assets in a recursive manner to refine the clustering solution further and create a more robust portfolio allocation strategy. After the initial hierarchical clustering step groups stocks based on their similarities, recursive bisection helps break down the resulting clusters into smaller sub-clusters, allowing for a more detailed and granular allocation of stocks within each cluster. Furthermore, by creating sub-clusters, recursive bisection reduces the number of stocks competing for weight allocation, meaning that assets within a sub-cluster will be more closely related, and their performance is likely to be more aligned. As a result, the allocation decision becomes less complex and may lead to more stable and reliable portfolio performance.
\subsubsection{Quasi-Diagonalization}
Once all stocks are hierarchically clustered and sub-clustered, we move on to performing quasi-diagonalization to our algorithm to build our covariance matrix and enhance the interpretation of the clustering results. I initially clustered all the stocks into a hierarchical tree based on previously defined similarity; however, I now rearranged the rows and columns of the covariance matrix of stocks so that similar stocks are placed together, and the stocks displaying the most variance are placed further apart. This quasi-diagonalization approach should now rearrange the covariance matrix so that the more significant covariances in the matrix are placed across the diagonal, with the smaller ones spread around the diagonal. 
\subsubsection{Building the Portfolio}
Our last step is to build the portfolio. In this analysis, I implement a long-short portfolio which allows me to take advantage of both upward and downward market movements. By incorporating short positions, I can analyze declining stock prices, thus enhancing the overall risk-adjusted returns and offering a more comprehensive strategy to capture market inefficiencies. To do this, I added a side\_weights parameter to signify stocks intended to be short versus long. For example, a -1 indicates going short on a stock, and a 1 indicates going long.
\section{Empirical Tests and Results}
By using a hierarchical clustering-based approach, I find that I can achieve improved out-of-sample portfolio performance on a risk-adjusted basis compared to the Markowitz Model by comparing them across various market conditions.
\FloatBarrier
\begin{figure}[!htb]
	\centering
	\includegraphics[width=\columnwidth]{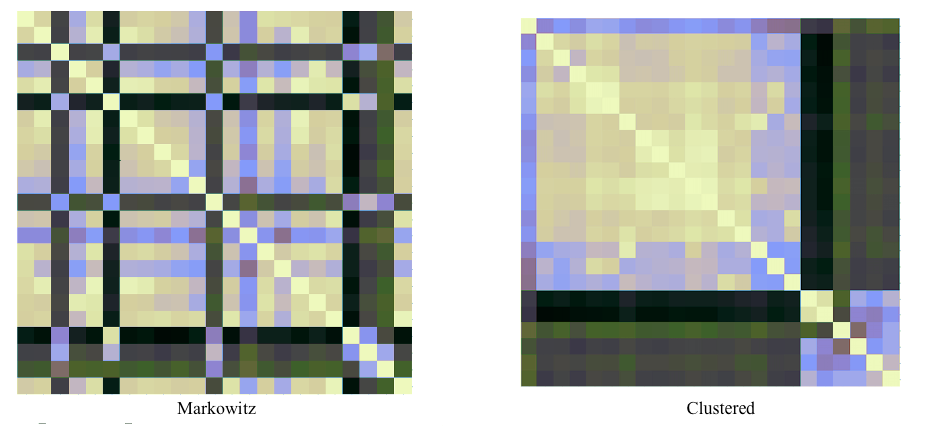}
    \captionsetup{labelformat=empty}
	\caption{Fig. 2: Stock Example}
         \subcaption{This figure plots the covariance matrix for the devised hierarchical clustering model and the covariance matrix for the traditional Markowitz model. The covariance matrices are heat-map colored by the correlation of the stocks. It is important to note that closely clustered stocks are grouped closer to the diagonal line.}
	\label{fig: Figure 2}
\end{figure}
\FloatBarrier
Figure 2 represents an illustrated example of where randomly selected stocks were used to depict correlation differences. Through the covariance matrices, we can see that the hierarchical clustering matrix can heat-map the correlation of the stocks much better than the Markowitz model can, as seen based on the color correlations. For example, on the clustered matrix, we can see that similar colored stocks are seen closer to each other on the hierarchy, indicating that similar stocks are being clustered together. Conversely, the stocks are less correlated on the Markowitz matrix, indicating incorrect allocations. Therefore, it is evident that the clustered portfolio groups stock better than the Markowitz Model does, allowing for a more robust correlation in a covariance matrix and a more diversified portfolio.
\FloatBarrier
\begin{figure}[!htb]
	\centering
	\includegraphics[width=\columnwidth]{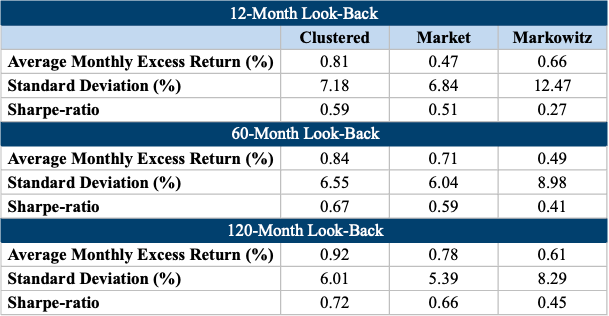}
     \captionsetup{labelformat=empty}
	\caption{Table 1: Portfolio Performance Analysis}
         \subcaption{In this study, portfolios were analyzed using a 3-month buy-and-hold, long-short strategy, using the side\_weights parameter. The following analysis was conducted using CRSP monthly data from 1965-2022. Using a 12-month, 60-month, and 120-month look-back period\cite{3b09d2c1-d82c-3d11-9b79-a42cdddc4257}, the analysis is based on out-of-sample return performance. To conduct my analysis of the portfolio performance, I look at the following three measures: Average Monthly Excess Returns, Standard Deviation, and the Sharpe-Ratios}
         \label{fig: Table 1}
\end{figure}
\FloatBarrier
Average monthly excess returns, standard deviation, and Sharpe-ratios are commonly used metrics to determine a portfolio allocation model's efficacy because they provide valuable insights into the risk and return characteristics of the portfolio \cite{article}. For starters, excess returns refer to the returns achieved by a portfolio above the risk-free rate or a benchmark index and measures the portfolio's performance relative to a baseline. By calculating the average monthly excess returns, investors can assess how well the portfolio is performing compared to the risk-free rate or a benchmark. Standard deviation, on the other hand, is a measure of the dispersion of returns around the portfolio's average return that quantifies the volatility or risk of the portfolio. A higher standard deviation indicates greater price fluctuation, implying higher risk. Finally, the Sharpe-ratio is a risk-adjusted performance metric that measures the amount of return per unit of risk taken, and enables investors to compare different portfolios' risk-adjusted returns and choose the one that offers the best trade-off between risk and return. Thus, Table 1 depicts the average monthly excess returns, standard deviation, and Sharpe-ratios for the clustered portfolio,
the Markowitz portfolio, and the standard market portfolio. As evident by the results, when looking at a 12-month, 60-month, and 120-month look-back period, the clustered portfolio merits better measures when compared to the Markowitz Model. Comparing these values to the market, we can see that the clustered portfolio can produce higher risk-adjusted returns and Sharpe-ratio measures by roughly 0.34\% and 0.08\%, respectively, while maintaining comparable standard deviation measures, thus indicating improved portfolio performance. The same, however, cannot be said for the Markowitz portfolio, which has a lower average monthly excess return rate and lower Sharpe-ratios, with a difference of 0.15\% and 0.32\%, respectively, while having significantly higher standard deviation rates of 5.29\%, when compared to the clustered portfolio. While it is beneficial for our model to be analyzed using recent data that mimics the trends seen in the stock market more closely, I also aimed to see how the model fares when being compared over a more extended look-back period, and thus, studied the portfolio performance of both models when using a 60-month and 120-month look-back window to better approximate the covariance matrix. With the 60-month and 120-month look-back windows, we can see that the trend continues where the clustered portfolio produces average monthly excess returns, standard deviation, and Sharpe-ratios that outperform the market value while notably exceeding the values produced by the Markowitz portfolio.
\FloatBarrier
\begin{figure}[!htb]
	\centering
	\includegraphics[width=\columnwidth]{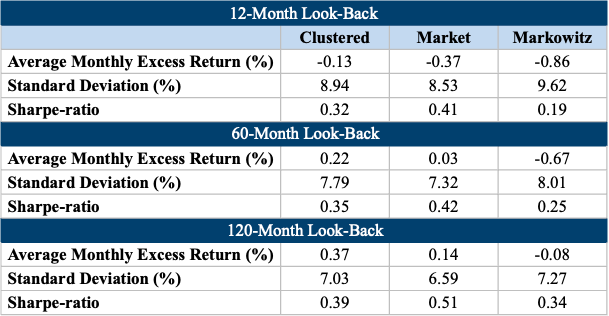}
    \captionsetup{labelformat=empty}
	\caption{Table 2: Portfolio Performance Analysis During Market Downturns}
         \subcaption{In this study, portfolios were analyzed using a 3-month buy-and-hold, long-short strategy, using the side\_weights parameter. The following analysis was conducted using CRSP monthly data from 1965-2022, but only using data from market downturns. Using 12-month, 60-month, and 120-month look-back periods, the analysis is based on out-of-sample return performance. To conduct my analysis of the portfolio performance, I look at the following three measures: Average Monthly Excess Returns, Standard Deviation, and the Sharpe-Ratios.}
	\label{fig: Table 2}
\end{figure}
\FloatBarrier
To further ensure the accuracy and usability of this model, I compared the clustered model and the Markowitz model using specifically a time series of returns from market downturns while looking at the same metrics seen in Table 1 of Average monthly excess returns, standard deviation, and Sharpe-ratios. Market downturns refer to periods of significant decline in the overall value of financial markets, which can have a profound impact on investors' portfolios, and thus are important to consider when evaluating models. As seen in Table 2, the clustered model relays the same trends as in Table 1, where the clustered model performs better than the Markowitz model along those same measures. When looking at the 12-month look-back period, precisely that of the clustered portfolio and the market, we can see that the clustered portfolio can produce higher risk-adjust returns by roughly 0.24\% while maintaining comparable standard deviation measures and Sharpe-ratios, thus indicating improved portfolio performance. On the other hand, the Markowitz portfolio has a lower average monthly excess returns and lower Sharpe-ratios, with a difference of 0.73\% and 0.13, respectively, while having significantly higher standard deviation rates of 0.68\%, when compared to the clustered portfolio. I similarly studied the portfolio performance of both models when using a 60-month and 120-month look-back window to better approximate the covariance matrix. With the 60-month and 120-month look-back windows, we can see that the trend continues where the clustered portfolio produces average monthly excess return rates, standard deviation, and Sharpe-ratios that outperform the market value while notably exceeding the values produced by the Markowitz portfolio. Therefore, on a risk-adjusted metric, the clustered portfolio has better excess returns during market downturns.  
\section{Conclusion}
Through this study, I was able to find significant and robust evidence of improved portfolio performance using a hierarchical clustering approach compared to the standard Markowitz approach. Using out-of-sample portfolio performance, studied using a large cross-section of U.S. traded stocks over the past sixty years, the clustered model shows improvements in portfolio allocation, which is vitally important to investors worldwide. By ensuring that results are not spurious, through the improvements in risk volatility and portfolio performance, I find a significant improvement in result consistency in our model compared to that of a Markowitz approach, which is of paramount importance when using real money in real time. The empirical tests and analyses conducted in this research demonstrate that the hierarchical clustering model outperforms the traditional Markowitz model across various market conditions. Notably, the clustered portfolio consistently yields higher risk-adjusted returns, better standard deviation rates, and improved Sharpe-ratios, even during market downturns, across several look-back periods, thus reinforcing the robustness and effectiveness of the approach. Furthermore, by incorporating a long-short strategy, the model allows investors to capitalize on both upward and downward market movements, offering a comprehensive strategy to exploit market inefficiencies.

\section{Implications}
This research contributes to financial economics and machine learning literature by providing an analysis of portfolio performance when using a clustered covariance matrix versus a traditional Markowitz covariance matrix, thereby improving asset allocation and portfolio optimization. As investors face the constant challenge of dividing their investments among various assets to optimize their portfolios, the need for improved computational methods has become increasingly evident. By utilizing machine learning in the asset allocation domain, this study presents an interesting approach to using hierarchical clustering to enhance optimization accuracy by introducing other model constraints to test the model's validity. 

By shedding light on the effectiveness of the hierarchical clustering technique in portfolio optimization, this study improves our understanding of portfolio performance and predictions about market behavior and offers valuable insights to investors worldwide. The findings pave the way for developing more sophisticated and accurate numerical models in the future, supporting practitioners and academics in making better money allocation decisions, especially in today's dynamic and complex financial landscape.
\section{Future Work}
In the future, I aim to test my model across various other models used in the industry and other risk management techniques. While the Markowitz model is the most common framework investors use to allocate their portfolios, I aim to test the clustered approach against similar models, such as the Black-Litterman Model and other Bayesian-based approaches. Similarly, evaluating the effectiveness of risk management techniques, such as using hedging strategies in combination with the hierarchical clustering approach, could be an exciting avenue for future research as it would provide insights into enhancing risk-adjusted returns and minimizing downside risk during market downturns.

Furthermore, I also plan to test alternate portfolio constraints. In this study, I used a long-short strategy for portfolio allocation, however, in the future, I aim to explore the impact of different constraints, such as maximum and minimum allocation limits for individual assets or sectors, and how these constraints affect portfolio performance. Additionally, while this study considers factors like size, value, profitability, investment, and momentum, future research could explore other variables or macroeconomic indicators that influence asset returns that can lead to more accurate and robust portfolio allocation strategies. 

Finally, I also look towards dynamic portfolio optimization, where allocation strategies adapt based on changing market conditions. The current study focuses on out-of-sample performance using a fixed look-back period. However, incorporating real-time data and dynamically adjusting portfolio weights could lead to more adaptive and responsive investment strategies. 
\section{Acknowledgment}
I would like to thank Dr. Stephen Owen\cite{Owen_2019} at the University of North Texas for his kind and thoughtful guidance and mentorship throughout this research project.
\bibliographystyle{IEEEtran}

\bibliography{Bibliography}

\end{document}